# Assessment of Efficiency, Impact Factor, Impact of Probe Mass, Probe Life Expectancy, and Reliability of Mars Missions

Malaya Kumar Biswal M[*] and Ramesh Naidu Annavarapu[†]

*Department of Physics*
*School of Physical Chemical and Applied Sciences,*
*Pondicherry University, Kalapet, Puducherry, India – 605 014*

Mars is the next frontier after Moon for space explorers to demonstrate the extent of human expedition and technology beyond low-earth orbit. Government space agencies as well as private space sectors are extensively endeavouring for a better space enterprise. Focusing on the inspiration to reach Mars by robotic satellite, we have interpreted some of the significant mission parameters like proportionality of mission attempts, efficiency and reliability of Mars probes, Impact and Impact Factor of mass on mission duration, Time lag between consecutive mission attempts, interpretation of probe life and transitional region employing various mathematical analysis. And we have discussed the importance of these parameters for a prospective mission accomplishment. Our novelty in this paper is we have found a deep relation describing that the probe mass adversely affects the mission duration. Applying this relation, we also interpreted the duration of probe life expectancy for upcoming missions.

## I. Nomenclature

| | | |
|---|---|---|
| *CNSA* | = | China National Space Administration |
| *ESA* | = | European Space Agency |
| *ISRO* | = | Indian Space Research Organization |
| *JAXA* | = | Japan Aerospace Exploration Agency |
| *MBSRC* | = | Mohammed Bin Rashid Space Center |
| *MGRSO* | = | Mars Global Remote Sensing Orbiter |
| *MMX* | = | Martian Moon Exploration |
| *MOM* | = | Mars Orbiter Mission |
| *NASA* | = | National Aeronautics Space Administration |
| *NICT* | = | National Institute of Information and Technology |

## II. Introduction

Journey to Mars has fascinated many enthusiasts and space scientists for planetary exploration. For a prosperous strive, it is significant to have a perspective knowledge of mission trends and their effects on the community and mankind. Henceforth with reference to our preceding paper [1, 2] we have assessed data and employed a mathematical regression analysis technique to interpret various mission parameters described in the abstract. These parameters are significant enough to determine mission prospects. Our assessment report is novel and unique and has found nowhere in any of the analysis reports. Thus these findings may have a potential impact on upcoming missions.

---

[*] Graduate Researcher, Department of Physics, Pondicherry University, India; **malaykumar1997@gmail.com**, **mkumar97.res@pondiuni.edu.in**, Member of Indian Science Congress Association, Student Member AIAA
[†] Associate Professor, Department of Physics, Pondicherry University, India; **rameshnaidu.phy@pondiuni.edu.in**, **arameshnaidu@gmail.com**.

## III. Terms, Definitions and Research Methodology

### A. Terms and Definitions

- Mission Duration: It is considered as the number of days from the date of launch to the date of the last operation (last contact).
- Mission Degradation: It is considered between the date of launch and to the date of decay (mission lost).

### B. Research Methodology

For our analysis, we gathered data from the dataset [3, 4] for the probe mass, mission duration/degradation. Similarly, the time lag is estimated between their consecutive launch and decay dates. As the time lag data shows a good response to our analysis we have taken the negative value data in positive. Additionally, the duration for the operational probe is considered from the date of launch to the date of operation as of 1st May 2020 of this current calendar year. The data gathered were plotted against the period from the 1960s to the 2020s and various mathematical techniques (Linear, Logarithmic, and Polynomial Regression Analysis) were performed to show predictive trend curves for all parameters. We also have interpreted the probe life expectancy curve that is capable of determining the lifespan of upcoming probes with respect to their mass. Because the probe mass has a great impact on its duration. We will discuss it further. Our assessment report is very different and novel and has not explained in any published reports or online resources. Hence we consider that this will provide an outline for a perspective idea for attempting successful missions in the near future.

## IV. Assessment of Mission Parameters

### A. Mission Attempt Rate

In this section, we have discussed the rate of mission attempts at the frequency of 20 years from the 1960s to the 2020s. These rates are the measure of the ratio of the number of attempts (success or failure) to the total number of attempts. Polynomial regression analysis and curve fitting method were executed against these ratios to obtain fine curves shown in Fig.1. It shows that the frequency of attempts that remained failure decreases from a higher peak to the lower level. Contradictory to this curve the frequency of success rises from the 1960s to the 2020s. And we found that the nation's economic standard, technological feasibility, and the rate of success and failure determine the mission attempt [5, 6]

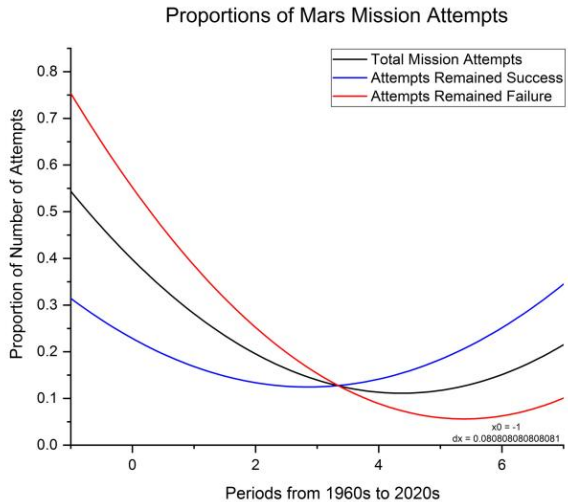
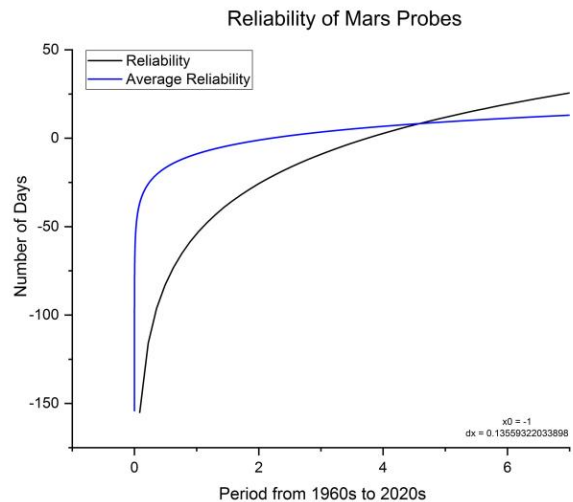

**Figure 1 Proportions of Mars Mission Attempts**     **Figure 2 Reliability of Mars Probes**

### B. Reliability

The reliability of the mission determines the longevity (the ability of a probe to withstand in orbit or planetary surface for operation) of Mars probes. It also determines the possibility of mission accomplishment to a greater extent. And we have estimated the difference between the period of degradation and duration as the reliability of the probe. Furthermore, its average is considered to be the average mission reliability. Performing the logarithmic regression analyzing method over the data acquired from [3] to obtain the reliability curve shown in Fig.2. The line of reliability shows an excellent improvement of probe reliability over the years. Inadequate fabrication and ground testing, space environmental condition, and robustness nature of the probe components are the considerable factors affecting the reliability of Mars probes. Mathematical expression for reliability is

$$Reliability = \frac{Degradation\ Period\ (Days)\ -\ Duration\ Period\ (Days)}{2}$$

### C. Impact Factor of Mass on Duration

Impact Factor is the measure of the ratio of the sum of duration/degradation of probes of two preceding attempts to the sum of masses of probes of two preceding attempts and mathematically expresses as

$$I.F_y = \frac{(Duration/Degradation)_{y-1} + (Duration/Degradation)_{y-2}}{Mass_{y-1} + Mass_{y-2}}$$

It displays the impact factor of mass on the duration/degradation of probes shown in Fig.3. Over the past 60 years. And we notice that there is a gradual increase in impact factor over duration contradictory to degradation. Hence, it indirectly shows the rise of mission duration over the years from the 1960s to the 2020s.

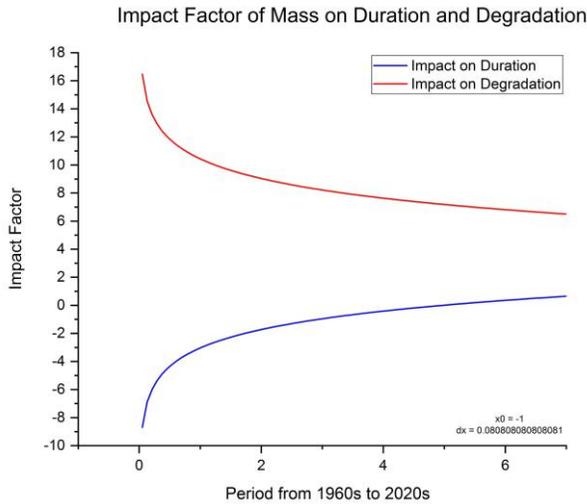
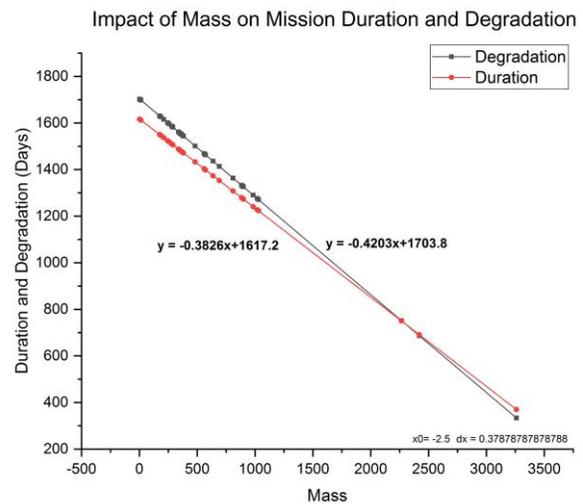

**Figure 3 Impact Factor of Mass on Duration and Degradation**   **Figure 4 Impact of Mass on Duration and Degradation**

### D. Impact of Mass on Mission Duration and Degradation

One of the significant relationships we found in this paper is the impact of mass on mission duration. It showed that the probe having lower masses have greater duration than the heavier probes. So, we performed linear regression and curve-fitting techniques for both duration and degradation with respect to their masses. It yielded two straight lines shown in Fig.4. The dense point along the highest point is the approximation of probe mass and both duration and degradation for the next 10 missions.

### E. Probe Life Expectancy

Concerning the section (4.4), we are interested to interpret the lifespan of the probes with reference to their masses. In this analysis, we have eliminated the data for the duration and degradation of the probe lost with launch vehicle issues in order to have good precision. And we used logarithmic regression analysis for the selected data gathered from [3, 4] to get two curves of probe life expectancy shown in Fig.5. Using these results, we have interpreted the lifespan of upcoming probes mentioned in table-1 and graphically shown in Fig.6. Similar to this we can also interpret lifespan using the line equation shown in Fig.4

**Table-1. Interpreted Mission Life Expectancy**

| Period from 2020 - 2024 Mission Name Equation Units | Agency | Mass Kg | Interpreted Results | | | Interpreted Results | | |
|---|---|---|---|---|---|---|---|---|
| | | | Duration Line Days | Duration Logarithmic Days | Average Days | Degradation Line Days | Degradation Logarithmic Days | Average Days |
| Mars 2020 Rover | NASA | 1025 | 1113.38 | 1225.04 | 1169.21 | 1152.58 | 1272.99 | 1212.79 |
| MGRSO/Tianwen-1 | CNSA | 3175 | 921.29 | 402.45 | 661.87 | 941.95 | 369.35 | 655.65 |
| MGRSO Rover | CNSA | 240 | 1360.04 | 1525.38 | 1442.71 | 1423.06 | 1602.93 | 1512.99 |
| Hope Mars Mission | MBRSC | 1500 | 1048.68 | 1043.30 | 1045.99 | 1081.65 | 1073.35 | 1077.50 |
| ExoMars 2020 Rover | ESA | 310 | 1316.56 | 1498.59 | 1407.58 | 1375.38 | 1573.51 | 1474.44 |
| Terahertz Explorer | NICT | 140 | 1451.61 | 1563.64 | 1507.63 | 1523.47 | 1644.96 | 1584.22 |
| Mangalyaan 2 / MOM-2 | ISRO | 100 | 1508.78 | 1578.94 | 1543.86 | 1586.16 | 1661.77 | 1623.96 |
| Martian Moon Exploration | JAXA | 150 | 1439.89 | 1559.81 | 1499.85 | 1510.62 | 1640.76 | 1575.69 |

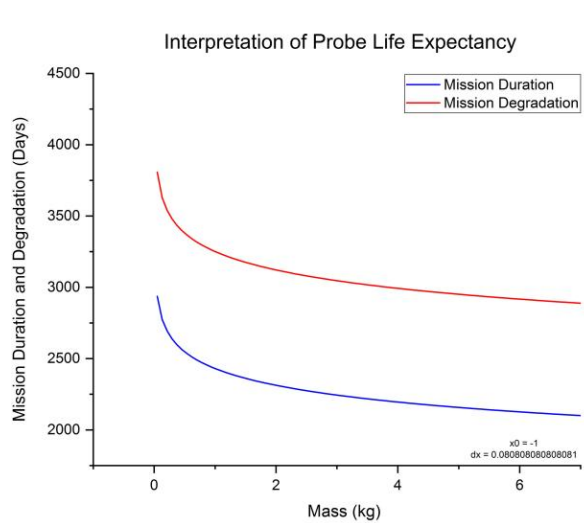
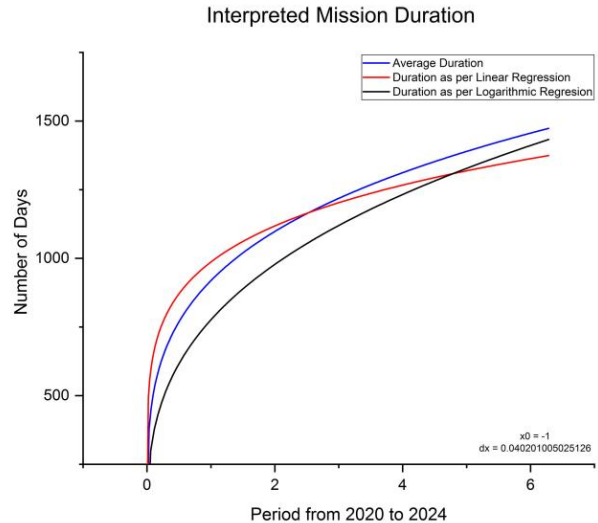

Figure 5 Probe Life Interpretation    Figure 6 Interpreted Mission Duration

## F. Efficiency of the Missions

The efficiency of the mission is estimated as the ratio of output (no. of days in operation) to the input (no. of days taken to prepare and launch subsequent probes) multiplied by 100.

$$Efficiency\ (\%) = \frac{Output\ (No.of.Days\ in\ Operation)}{Input\ (No.of.Days\ taken\ for\ concurrent\ launch)} \times 100$$

From the ratio described in the dataset [4], we have applied the logarithmic regression technique to determine efficiency in three epochs shown in Fig.7. We observe that the period (the 1960s-1980s, and 2000s – 2020s) have greater efficiency than the period (1980s – 2000s). It may be due to repeated or number of failures during that epoch.

## G. Mission Intermission

Additional to the other parameters, we also have estimated mission intermission curves between two consecutive launch and decay date intervals (in terms of the number of days). Performing the logarithmic regression analysis method, we obtained two fine curves shown in Fig.8. We describe that the input effort in launching the probes and their degradation intervals are almost the same as the curves go parallel to each other. Further, these curves closely explain the gradual increase in mission efficiency over the years.

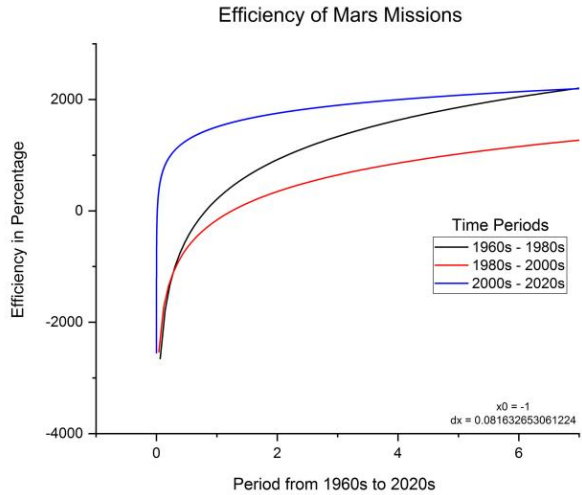
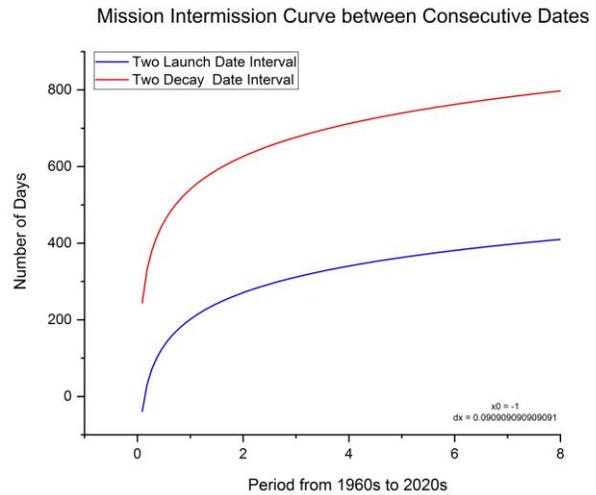

Figure 7 Efficiency of Mars Missions        Figure 8 Mission Intermission Curves

## H. Transitional and Active Region

The transitional region or intermediate region is the region of space lies between the curves of mission duration and degradation shown in Fig.9. In this region, the probes start to decay after either accomplishing its mission target or losing its function. Similarly, the region below the duration curve is the active region where the probe starts its mission, accomplishes the goal, shows greater performance, and the incapability of functioning well.

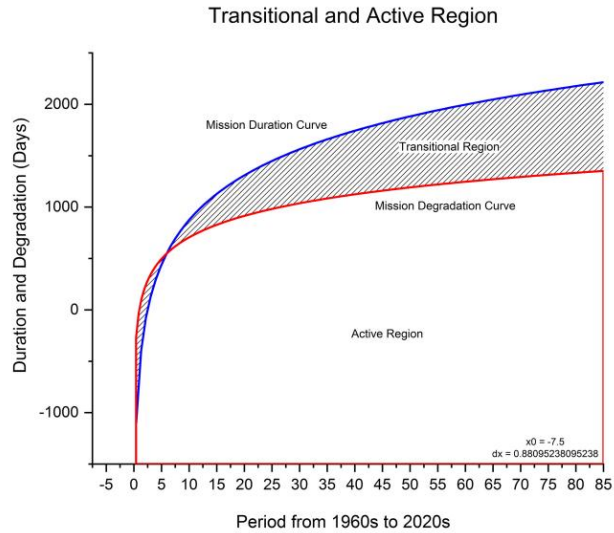

**Figure 9 Transitional and Active Region**

## V. Results and Discussions

**Table-2. Resultant Equations of all Mission Parameters**

| Eq. No | Equation Name | Equations | $R^2$ Value | Figure Reference |
|---|---|---|---|---|
| | | **Proportion of Mission Attempts** | | |
| 01. | Total Attempts | $y = 0.015(x^2) - 0.1311(x) + 0.3977$ | 0.5815 | Figure-1 |
| 02. | Failed Attempts | $y = 0.0171(x^2) - 0.1842(x) + 0.5522$ | 0.8647 | Figure-1 |
| 03. | Success Attempts | $y = 0.0128(x^2) - 0.0730(x) + 0.2286$ | 0.2368 | Figure-1 |
| | | **Reliability** | | |
| 04. | Reliability | $y = 40.929 \ln(x) - 54.042$ | 0.0644 | Figure-2 |
| 05. | Average Reliability | $y = 11.258 \ln(x) - 8.8752$ | 0.0224 | Figure-2 |
| | | **Impact Factor of Mass on Duration and Degradation** | | |
| 06. | Duration | $y = 1.893 \ln(x) - 3.0375$ | 0.1496 | Figure-3 |
| 07. | Degradation | $y = -2.021 \ln(x) + 10.437$ | 0.0057 | Figure-3 |
| | | **Impact of Mass on Duration and Degradation** | | |
| 08. | Duration | $y = -0.3826(x) + 1617.2$ | 0.0477 | Figure-4 |
| 09. | Degradation | $y = -0.4203(x) + 1703.8$ | 0.0570 | Figure-4 |
| | | **Probe Life Interpretation** | | |
| 10. | Duration | $y = -169.6 \ln(x) + 2291.2$ | 0.538 | Figure-5 |
| 11. | Degradation | $y = -186.3 \ln(x) + 2444.1$ | 0.538 | Figure-5 |
| | | **Mission Efficiency** | | |
| 12. | 1960s – 1980s | $y = 1024.6 \ln(x) + 209.46$ | 0.0635 | Figure-6 |
| 13. | 1980s – 2000s | $y = 736.14 \ln(x) - 163.36$ | 0.0942 | Figure-6 |
| 14. | 2000s – 2020s | $y = 352.65 \ln(x) + 1508.6$ | 0.0063 | Figure-6 |
| | | **Mission Intermission between Two Consecutive Dates** | | |
| 15. | Launch Date Interval | $y = 100.26 \ln(x) + 201.41$ | 0.0100 | Figure-7 |
| 16. | Decay Date Interval | $y = 123.52 \ln(x) + 540.84$ | 0.0094 | Figure-7 |
| | | **Transitional Region** | | |
| 17. | Duration Curve | $y = 626.47 \ln(x) - 568.35$ | 0.1035 | Figure-8 |
| 18. | Degradation Curve | $y = 302.1 \ln(x) + 9.5611$ | 0.0541 | Figure-8 |
| | | **Interpreted Mission Duration** | | |
| 19. | Duration as per Eq. (8) | $y = 986.62(x)^{0.1803}$ | 0.5006 | Figure-9 |
| 20. | Duration as per Eq. (10) | $y = 776.24(x)^{0.3335}$ | 0.2511 | Figure-9 |
| 21. | Average Duration | $y = 918.76(x)^{0.2570}$ | 0.3400 | Figure-9 |

Performing various mathematical interpretation techniques, we obtained 2 line, 3 power, 3 polynomials, and 13 logarithmic regression equations shown in table-2. And its plotted graphs were shown in appropriate sections. From overall observations, the relationship of the impact factor and impact of mass on mission duration greatly promises for a durable mission. The interpreted duration for future missions from this analysis is uncertain in data accuracy. However, we can roughly approximate the duration. The essential step is to reduce the probe mass and the probe masses ranging between 200 kilograms to 1000 kilograms suit best for the mission reliability. Then coming to the transitional region graph shown in Fig.8, the degradation line is supposed to be above the duration curve, it is because the resultant curve obtained shows the progression of mission duration over years from the regression analysis. So, it ultimately suppresses the degradation curve to lie under the duration curve. Further the efficiency we estimated as per the acquired data is tentative and can vary from probe to probe. Because the exact efficiency can be counted as per the data and results returned by the probe either from orbit or the planetary surface (On Mars) to the ground (On Earth). Moreover, the reliability of the mission is assessed based on the number of degradation and duration. It is supposed to be considered depending on the rate of physical tolerance (space environmental factors) and the internal maintenance of components (efficient feasibility of technological components and circuitries).

## VI. Conclusions

Concerning the mission tragedies and future prospects, we have clearly explained the mission parameters and interpreted various results by various mathematical interpretation techniques. The results and interpretations were graphically shown in each appropriate section. The resultant equations of mission parameters were orderly showed in table-2. Hence, we conclude that despite the uncertainty of some of the data in this analysis, the trend of Mars mission parameters and interpretation will greatly help global space communities to begin with a gait leap towards planetary explorations.

**About the Work**

The work has been carried out during the Covid-19 Pandemic and subjected to peer review process. The terms and definitions are novel and nowhere found elsewhere during the search for literature studies. This study was performed at the Department of Physics, Pondicherry University to show and analyze various mission parameters and its effects on mission life employing mathematical techniques such as Regression Analysis, Extrapolation and Averaging.

**Copyright License and Statement**

**Dedication**

The main author Malaya Kumar Biswal M would like to dedicate this work to his beloved mother late. Mrs Malathi Biswal for her motivational speech and emotional support throughout his life.

**Supplementary Reading**

**Mars Missions Failure Report Assortment: Review and Conspectus.** Presented as Technical Paper at 2020 AIAA Propulsion and Energy Forum on 24 August 202 with paper number AIAA-2020-3541.Published by American Institute of Aeronautics and Astronautics. **https://arc.aiaa.org/doi/abs/10.2514/6.2020-3541.**

# Appendices

## Data for Assessment of Durability, Efficiency, Impact Factor, Impact of Probe Mass on mission duration, and Probe Life Expectancy of Mars Missions

### Table-3 Overall Mission Attempt Rate

| Period | Number of Counts | | | Proportions | | |
|---|---|---|---|---|---|---|
| | Total Attempts | Success Attempts | Failed Attempts | Total Attempts | Success Attempts | Failed Attempts |
| 1960s-1970s | 12 | 3 | 9 | 0.272727273 | 0.142857143 | 0.391304348 |
| 1970s-1980s | 11 | 5 | 6 | 0.25 | 0.238095238 | 0.260869565 |
| 1980s-1990s | 2 | 0 | 2 | 0.045454545 | 0 | 0.086956522 |
| 1990s-2000s | 7 | 3 | 4 | 0.159090909 | 0.142857143 | 0.173913043 |
| 2000s-2010s | 6 | 5.5 | 0.5 | 0.136363636 | 0.261904762 | 0.02173913 |
| 2010s-2020s | 6 | 4.5 | 1.5 | 0.136363636 | 0.214285714 | 0.065217391 |
| **Total** | **44** | **21** | **23** | | | |

### Table-4 Efficiency of Mars Missions

| Period | 1960s-1980s | | | 1980s-2000s | | | 2000s-2020s | | |
|---|---|---|---|---|---|---|---|---|---|
| Counts | Output† | Input§ | Efficiency | Output† | Input§ | Efficiency | Output† | Input§ | Efficiency |
| 1 | 0.0034 | 4 | 0.085 | 52 | 4685 | 0.085 | 6964 | 825 | 0.085 |
| 2 | 0.0034 | 740 | 0.000459 | 52 | 5 | 0.000459 | **6178** | | |
| 3 | 140 | 8 | 1750 | 258 | | - | 206 | 786 | 1750 |
| 4 | 0.003 | 3 | 0.1 | 330 | 1536 | 0.1 | 2477 | 8 | 0.1 |
| 5 | 0.36 | 732 | 0.04918 | 3647 | 1536 | 0.04918 | 5452 | 8 | 0.04918 |
| 6 | 1118 | 23 | 4860.87 | 0.0062 | 1504 | 4860.87 | **5376** | 28 | 4860.87 |
| 7 | 249 | 2 | 12450 | - | 9 | - | 456 | 766 | 12450 |
| 8 | 666 | 1548 | 43.02326 | - | 18 | - | 0.00155 | 722 | 43.02326 |
| 9 | 0.005 | 30 | 0.016667 | 297 | - | | - | 1557 | - |
| 10 | 640 | 30 | 2133.333 | 219 | - | | - | 18 | - |
| 11 | 0.00048 | 6 | 0.008 | 1983 | 576 | 0.008 | **3079** | - | - |
| 12 | 0.0032 | 767 | 0.000417 | 286 | 576 | 0.000417 | **2369** | - | - |
| 13 | 0.0625 | 1 | 6.25 | 334 | 161 | 6.25 | **2356** | 710 | 6.25 |
| 14 | 461 | 9 | 5122.222 | 334 | 23 | 5122.222 | **1509** | 13 | 5122.222 |
| 15 | 192 | 9 | 2133.333 | | | | 219 | 847 | 2133.333 |
| 16 | - | | | | | | **727** | 71 | |
| 17 | 452 | | | | | | 244 | 71 | |
| 18 | 188 | 2 | 9400 | | | | 238 | | |
| 19 | - | 2 | | | | | | | |
| 20 | 514 | | | | | | | | |
| 21 | 195 | 783 | 24.90421 | | | | | | |
| 22 | 218 | 4 | 5450 | | | | | | |
| 23 | 219 | 11 | 1990.909 | | | | | | |
| 24 | 219 | 4 | 5475 | | | | | | |
| 25 | 214 | | | | | | | | |
| 26 | 214 | 741 | 28.87989 | | | | | | |
| 27 | 1824 | 741 | 246.1538 | | | | | | |
| 28 | 2040 | 20 | 10200 | | | | | | |
| 29 | 1050 | 20 | 5250 | | | | | | |
| 30 | 1677 | 4685 | 35.79509 | | | | | | |
| **Total** | **12490.44** | **10925** | **66600.93** | **7792.006** | **10629** | **9989.585** | **37850** | **6430** | **26365.93** |

**Notes:**
- † **Output** – Total number of durations (Days)
- § **Input** - Total number of time elapsed for next mission launch (Days)
- Bolded fonts are operational missions

**Table-5 Durability of Mars Missions**

| S.No Units | Duration Days | Degradation Days | Average. Reliability Days | S.No Units | Duration Days | Degradation Days | Average. Reliability Days |
|---|---|---|---|---|---|---|---|
| 1 | 0.0036 | 1 | 0.4982 | 34 | 258 | 258 | 0 |
| 2 | 0.0034 | 1 | 0.4983 | 35 | 330 | 340 | 5 |
| 3 | 0.0034 | 125 | 62.4983 | 36 | 3647 | 3700 | 26.5 |
| 4 | 140 | 230 | 45 | 37 | 0.0062 | 2 | 0.9969 |
| 5 | 0.003 | 227 | 113.4985 | 38 | - | 2 | 0 |
| 6 | 0.36 | 1 | 0.32 | 39 | - | 2 | 0 |
| 7 | 1118 | 1118 | 0 | 40 | 297 | 297 | 0 |
| 8 | 249 | 249 | 0 | 41 | 219 | 219 | 0 |
| 9 | 666 | 666 | 0 | 42 | 1983 | 1985 | 1 |
| 10 | 0.005 | 1 | 0.4975 | 43 | 286 | 288 | 1 |
| 11 | 640 | 641 | 0.5 | 44 | 334 | 379 | 22.5 |
| 12 | 0.00048 | 1 | 0.49976 | 45 | 334 | 379 | 22.5 |
| 13 | 0.0032 | 1 | 0.4984 | 46 | **6964** | - | 0 |
| 14 | 0.0625 | 2 | 0.96875 | 47 | **6178** | - | 0 |
| 15 | 461 | 461 | 0 | 48 | 206 | 244 | 19 |
| 16 | 192 | 192 | 0 | 49 | 2477 | 2906 | 214.5 |
| 17 | - | 192 | 0 | 50 | 5452 | 5700 | 124 |
| 18 | 452 | 452 | 0 | 51 | **5376** | - | 0 |
| 19 | 188 | 188 | 0 | 52 | 456 | 1024 | 284 |
| 20 | - | 188 | 0 | 53 | 0.00155 | 68 | 33.999225 |
| 21 | 514 | 516 | 1 | 54 | - | 68 | 0 |
| 22 | 195 | 204 | 4.5 | 55 | - | 68 | 0 |
| 23 | 218 | 218 | 0 | 56 | **3079** | - | 0 |
| 24 | 219 | 219 | 0 | 57 | **2369** | - | 0 |
| 25 | 219 | 219 | 0 | 58 | **2356** | - | 0 |
| 26 | 214 | 228 | 7 | 59 | **1509** | - | 0 |
| 27 | 214 | 212 | -1 | 60 | 219 | 219 | 0 |
| 28 | 1824 | 1846 | 11 | 61 | **727** | - | 0 |
| 29 | 2040 | 2640 | 300 | 62 | 244 | 638 | 197 |
| 30 | 1050 | 1050 | 0 | 63 | 238 | 638 | 200 |
| 31 | 1677 | 1677 | 0 | | | | |
| 32 | 52 | 56 | 2 | | | | |
| 33 | 52 | 56 | 2 | | | | |
| | | | | | | | |
| **Total** | 58132.45 | 33502 | 1703.77 | | | | |
| **Average** | 922.73 | 531.78 | 27.04 | | | | |

**Table-6 Impact Factor of Mass on Duration and Degradation of Mars Probes**

| S.No | Mass | Duration | Degradation | Impact Factor (Mass on Duration) | Impact Factor (Mass on Degradation) |
|---|---|---|---|---|---|
| Units | Kg | Days | Days | - | - |
| 01. | 480 | 0.0036 | 01 | 0.00000075 | 0.00208 |
| 02. | 480 | 0.0034 | 01 | 0.00000071 | 0.00208 |
| 03. | 893 | 0.0034 | 125 | 0.00000495 | 0.091 |
| 04. | 893 | 140 | 230 | 0.0785 | 0.198 |
| 05. | 890 | 0.0030 | 227 | 0.0785 | 0.256 |
| 06. | 260 | 0.36 | 01 | 0.00031 | 0.198 |
| 07. | 244 | 1118 | 1118 | 2.212 | 2.220 |
| 08. | 890 | 249 | 249 | 1.205 | 1.205 |
| 09. | 381 | 666 | 666 | 0.719 | 0.719 |
| 10. | 3800 | 0.0050 | 01 | 0.159 | 00159 |
| 11. | 381 | 640 | 641 | 0.153 | 0.153 |
| 12. | 3800 | 0.00048 | 01 | 0.153 | 0.153 |
| 13. | 558.8 | 0.0032 | 01 | 0.00000084 | 0.000458 |
| 14. | 4549 | 0.0625 | 02 | 0.000012 | 0.0000587 |
| 15. | 2628 | 653 | 653 | 0.090 | 0.091 |
| 16. | 2628 | 640 | 640 | 0.246 | 0.285 |
| 17. | 558.8 | 514 | 516 | 0.362 | 0.362 |
| 18. | 2265 | 195 | 204 | 0.251 | 0.254 |
| 19. | 2265 | 218 | 218 | 0.091 | 0.093 |
| 20. | 2535 | 438 | 438 | 0.136 | 0.136 |
| 21. | 2535 | 428 | 440 | 0.170 | 0.173 |
| 22. | 1455 | 4464 | 4486 | 1.226 | 1.234 |
| 23. | 1455 | 2727 | 2727 | 2.471 | 2.478 |
| 24. | 2990 | 104 | 112 | 0.636 | 0.638 |
| 25. | 2990 | 516 | 535 | 0.103 | 0.108 |
| 26. | 1018 | 330 | 340 | 0.211 | 0.218 |
| 27. | 1030 | 3647 | 3700 | 1.941 | 1.972 |
| 28. | 3975 | 0.0186 | 06 | 0.7286 | 0.740 |
| 29. | 211 | 516 | 516 | 0.123 | 0.124 |
| 30. | 258 | 1983 | 1985 | 5.328 | 5.332 |
| 31. | 358 | 286 | 288 | 3.683 | 3.689 |
| 32. | 292.4 | 334 | 389 | 0.953 | 1.025 |
| 33. | 376 | 6964 | - | 10.918 | - |
| 34. | 646 | 6384 | 244 | 13.060 | - |
| 35. | 174 | 2477 | 2906 | 10.806 | 3.84 |
| 36. | 185 | 5452 | 5700 | 22.086 | 23.97 |
| 37. | 984 | 5376 | - | 9.262 | - |
| 38. | 350 | 456 | 1024 | 4.371 | - |
| 39. | 1781 | 0.0031 | 68 | 0.213 | 0.51 |
| 40. | 899 | 3079 | - | 1.148 | - |
| 41. | 482 | 2369 | - | 3.944 | - |
| 42. | 809 | 2356 | - | 3.659 | - |
| 43. | 690 | 1509 | - | 2.578 | - |
| 44. | 280 | 219 | 219 | 1.781 | - |
| 45. | 358 | 727 | - | 1.482 | - |
| 46. | 27 | 782 | 1276 | 3.140 | - |
| Total | 57988 | 58956.46 | 32894 | 111.9559293 | 211.4696767 |
| Average | 1260.60 | 1281.66 | 715.08 | 2.43 | 4.59 |

**Table-7 Predicted Lifespan from Regression Analysis**

| S.No | Duration | Mass | Degradation | Mass | Predicted Duration | Predicted Degradation |
|---|---|---|---|---|---|---|
| 1 | 140 | 893 | 230 | 893 | 1275.54 | 1328.55 |
| 2 | 1118 | 244 | 1118 | 244 | 1523.86 | 1601.30 |
| 3 | 249 | 890 | 249 | 890 | 1276.69 | 1329.81 |
| 4 | 666 | 381 | 666 | 381 | 1471.44 | 1543.72 |
| 5 | 640 | 381 | 641 | 381 | 1471.44 | 1543.72 |
| 6 | 461 | 2265 | 461 | 2265 | 750.59 | 751.96 |
| 7 | 192 | 362.5 | 192 | 362.5 | 1478.52 | 1551.50 |
| 8 | 452 | 2265 | 452 | 2265 | 750.59 | 751.96 |
| 9 | 188 | 362.5 | 188 | 362.5 | 1478.52 | 1551.50 |
| 10 | 514 | 558.8 | 516 | 558.8 | 1403.41 | 1469.00 |
| 11 | 195 | 2265 | 204 | 2265 | 750.59 | 751.96 |
| 12 | 218 | 2265 | 218 | 2265 | 750.59 | 751.96 |
| 13 | 219 | 3260 | 219 | 3260 | 369.88 | 333.80 |
| 14 | 219 | 3260 | 219 | 3260 | 369.88 | 333.80 |
| 15 | 214 | 3260 | 228 | 3260 | 369.88 | 333.80 |
| 16 | 214 | 3260 | 212 | 3260 | 369.88 | 333.80 |
| 17 | 1824 | 883 | 1846 | 883 | 1279.37 | 1332.75 |
| 18 | 2640 | 572 | 2640 | 572 | 1398.36 | 1463.45 |
| 19 | 1050 | 883 | 1050 | 883 | 1279.37 | 1332.75 |
| 20 | 1677 | 572 | 1677 | 572 | 1398.36 | 1463.45 |
| 21 | 52 | 2420 | 56 | 2420 | 691.28 | 686.82 |
| 22 | 52 | 570 | 56 | 570 | 1399.13 | 1464.29 |
| 23 | 258 | 2420 | 258 | 2420 | 691.28 | 686.82 |
| 24 | 258 | 570 | 277 | 570 | 1399.13 | 1464.29 |
| 25 | 330 | 1018 | 340 | 1018 | 1227.71 | 1276.02 |
| 26 | 3647 | 1030 | 3700 | 1030 | 1223.12 | 1270.97 |
| 27 | 297 | 210 | 297 | 210 | 1536.87 | 1615.59 |
| 28 | 219 | 11 | 219 | 11 | 1613.01 | 1699.22 |
| 29 | 1983 | 258 | 1985 | 258 | 1518.50 | 1595.41 |
| 30 | 286 | 338 | 288 | 338 | 1487.89 | 1561.79 |
| 31 | 334 | 290 | 379 | 290 | 1506.26 | 1581.97 |
| 32 | 334 | 2.4 | 379 | 2.4 | 1616.30 | 1702.83 |
| 33 | 6964 | 376 | | 376 | 1473.35 | 1545.82 |
| 34 | 6178 | 637 | | 637 | 1373.49 | 1436.14 |
| 35 | 206 | 9 | 244 | 9 | 1613.77 | 1700.06 |
| 36 | 2477 | 174 | 2906 | 174 | 1550.64 | 1630.71 |
| 37 | 5452 | 185 | 5700 | 185 | 1546.43 | 1626.09 |
| 38 | 5376 | 984 | | 984 | 1240.72 | 1290.31 |
| 39 | 456 | 350 | 1024 | 350 | 1483.30 | 1556.75 |
| 40 | 3079 | 899 | | 899 | 1273.24 | 1326.03 |
| 41 | 2369 | 482 | | 482 | 1432.80 | 1501.28 |
| 42 | 2356 | 809 | | 809 | 1307.68 | 1363.85 |
| 43 | 1509 | 690.8 | | 690.8 | 1352.91 | 1413.53 |
| 44 | 219 | 280 | 219 | 280 | 1510.08 | 1586.17 |
| 45 | 727 | 358 | | 358 | 1480.24 | 1553.39 |
| 46 | 244 | 13.5 | 638 | 13.5 | 1612.05 | 1698.17 |
| 47 | 238 | 13.5 | 638 | 13.5 | 1612.05 | 1698.17 |

**Table-8 Mission Intermission Between Two Consecutive Launch and Decay Date Interval**

| S.No | Launch Date Interval | Decay Date Interval | S.No | Launch Date Interval | Decay Date Interval |
|---|---|---|---|---|---|
| Units | Days | Days | Units | Days | Days |
| 1 | 4 | 4 | 33 | - | - |
| 2 | 740 | 865 | 34 | 1536 | 19 |
| 3 | 8 | 113 | 35 | 1536 | 1589 |
| 4 | 3 | 657 | 36 | 1504 | 4821 |
| 5 | 732 | 657 | 37 | 9 | 3636 |
| 6 | 23 | 1140 | 38 | 18 | 313 |
| 7 | 2 | 1098 | 39 | - | - |
| 8 | 1548 | 2196 | 40 | - | - |
| 9 | 30 | 636 | 41 | 576 | 78 |
| 10 | 30 | 641 | 42 | 576 | 2342 |
| 11 | 6 | 635 | 43 | 161 | 1538 |
| 12 | 767 | 767 | 44 | 23 | 71 |
| 13 | 1 | 3 | 45 | 825 | 1483 |
| 14 | 9 | 468 | 46 | - | - |
| 15 | 9 | 269 | 47 | 786 | - |
| 16 | - | - | 48 | 8 | - |
| 17 | - | - | 49 | 8 | 2279 |
| 18 | 2 | 264 | 50 | 28 | 3002 |
| 19 | 2 | 330 | 51 | 766 | 3507 |
| 20 | - | - | 52 | 722 | - |
| 21 | 783 | 471 | 53 | 1557 | 456 |
| 22 | 4 | 18 | 54 | 18 | 68 |
| 23 | 11 | 12 | 55 | - | - |
| 24 | 4 | 13 | 56 | - | - |
| 25 | - | - | 57 | 710 | - |
| 26 | 741 | 16 | 58 | 13 | - |
| 27 | 741 | 3169 | 59 | 847 | - |
| 28 | 20 | 2305 | 60 | 71 | - |
| 29 | 20 | 1362 | 61 | 71 | 219 |
| 30 | 4685 | 1317 | 62 | - | - |
| 31 | 4685 | 4444 | 63 | - | 244 |
| 32 | 5 | 144 | 64 | - | 603 |

**Table-9 Interpretation of Duration for future missions**

| Period from 2020 - 2024 | | Interpreted Results | | | Interpreted Results | | |
|---|---|---|---|---|---|---|---|
| Mission Name | Mass | Duration Line | Duration Logarithmic | Average | Degradation Line | Degradation Logarithmic | Average |
| Equation Units | Kg | Days | Days | Days | Days | Days | Days |
| Mars 2020 Rover | 1025 | 1113.38 | 1225.04 | 1169.21 | 1152.58 | 1272.99 | 1212.79 |
| Mars Global Remote Sensing Orbiter | 3175 | 921.29 | 402.45 | 661.87 | 941.95 | 369.35 | 655.65 |
| MGRSO Rover | 240 | 1360.04 | 1525.38 | 1442.71 | 1423.06 | 1602.93 | 1512.99 |
| Hope Mars Mission | 1500 | 1048.68 | 1043.30 | 1045.99 | 1081.65 | 1073.35 | 1077.50 |
| ExoMars 2020 Rover | 310 | 1316.56 | 1498.59 | 1407.58 | 1375.38 | 1573.51 | 1474.44 |
| Terahertz Explorer | 140 | 1451.61 | 1563.64 | 1507.63 | 1523.47 | 1644.96 | 1584.22 |
| Mangalyaan 2 | 100 | 1508.78 | 1578.94 | 1543.86 | 1586.16 | 1661.77 | 1623.96 |
| Martian Moon Exploration | 150 | 1439.89 | 1559.81 | 1499.85 | 1510.62 | 1640.76 | 1575.69 |

**Table-10 Selected Data for Mars Probe Life Span Prediction**

| Type | S.No | Spacecraft | Mass (kg) | Mission Duration** | Mission Degradation*** |
|---|---|---|---|---|---|
| Flybys | 01. | ~~1M.No.1~~ | ~~480~~ | ~~0.0036~~ | ~~01~~ |
| | 02. | ~~1M.No.2~~ | ~~480~~ | ~~0.0034~~ | ~~01~~ |
| | 03. | ~~2MV-4.No.1~~ | ~~893~~ | ~~0.0034~~ | ~~125~~ |
| | 04. | 2MV-4.No.2 | 893 | 140 | 230 |
| | 05. | ~~Mariner 3~~ | ~~260~~ | ~~0.36~~ | ~~01~~ |
| | 06. | Mariner 4 | 244 | 1118 | 1118 |
| | 07. | Zond 2 | 890 | 18 | 249 |
| | 08. | Mariner 6 | 381 | 670 | 666 |
| | 09. | Mariner 7 | 381 | 640 | 641 |
| | 10. | Mars 6 | 1900 | 219 | 219 |
| | 11. | Mars 7 | 1900 | 214 | 228 |
| | 12. | MarCo-A | 13.5 | 244 | 244 |
| | 13. | MarCo-B | 13.5 | 603 | 603 |
| Landers | 14. | ~~2MV-3.No.1~~ | ~~890~~ | ~~0.0030~~ | ~~01~~ |
| | 15. | Mars 2 | 358 | 192 | 192 |
| | 16. | Mars 3 | 358 | 188 | 188 |
| | 17. | Mars 6 | 635 | 219 | 219 |
| | 18. | Mars 7 | 635 | 214 | 228 |
| | 19. | Viking 1 | 572 | 2036 | 2036 |
| | 20. | Viking 2 | 572 | 1316 | 1316 |
| | 21. | Phobos 1 | 570 | 52 | 119 |
| | 22. | Phobos 2 | 570 | 258 | 258 |
| | 23. | ~~Mars 96~~ | ~~75~~ | ~~02~~ | ~~02~~ |
| | 24. | Mars Pathfinder | 210 | 297 | 360 |
| | 25. | Mars Polar Lander | 290 | 334 | 334 |
| | 26. | Beagle 2 | 09 | 183 | 206 |
| | 27. | Phoenix | 350 | 455 | 456 |
| | 28. | Schiaparelli EDM | 280 | 219 | 219 |
| | 29. | **InSight Mars Lander** | 358 | 727* | 727* |
| Orbiters | 30. | ~~2M.No.521~~ | ~~3800~~ | ~~0.0050~~ | ~~01~~ |
| | 31. | ~~2M.No.522~~ | ~~3800~~ | ~~0.00048~~ | ~~01~~ |
| | 32. | ~~Mariner 8~~ | ~~558.8~~ | ~~0.0032~~ | ~~01~~ |
| | 33. | ~~Kosmos 419~~ | ~~4549~~ | ~~0.0625~~ | ~~02~~ |
| | 34. | Mars 2 | 2265 | 461 | 461 |
| | 35. | Mars 3 | 2265 | 452 | 452 |
| | 36. | Mars 4 | 2265 | 195 | 204 |
| | 37. | Mars 5 | 2265 | 218 | 218 |
| | 38. | Viking 1 | 883 | 1846 | 2640 |
| | 39. | Viking 2 | 883 | 1050 | 1317 |
| | 40. | Phobos 1 | 2420 | 52 | 119 |
| | 41. | Phobos 2 | 2420 | 258 | 258 |
| | 42. | ~~Mars 96~~ | ~~3780~~ | ~~0.0062~~ | ~~02~~ |
| | 43. | Nozomi | 258 | 1983 | 1985 |
| | 44. | Mars Climate Orbiter | 338 | 286 | 286 |
| | 45. | **Mars Odyssey** | 376 | 6964* | 6964* |
| | 46. | **Mars Express** | 637 | 6178* | 6178* |
| | 47. | **Mars Reconnaissance Orbiter** | 984 | 5376* | 5376* |
| | 48. | ~~Fobos-Grunt~~ | ~~1560~~ | ~~0.00155~~ | ~~68~~ |
| | 49. | ~~Yinghuo-1~~ | ~~115~~ | ~~0.00155~~ | ~~68~~ |
| | 50. | **Mangalyaan** | 482 | 2369* | 2369* |
| | 51. | **MAVEN** | 809 | 2256* | 2256* |
| | 52. | **ExoMars TGO** | 690.8 | 1509* | 1509* |
| Rovers | 53. | Mars 2 – Prop-M | 4.5 | 192 | 192 |
| | 54. | Mars 3 – Prop-M | 4.5 | 188 | 188 |
| | 55. | Sojourner | 11 | 282 | 297 |
| | 56. | Spirit | 174 | 2269 | 2477 |
| | 57. | Opportunity | 185 | 5452 | 5700 |
| | 58. | **Curiosity** | 899 | 3079* | 3079* |

**Notes:**
- * Operational Mission – Duration Estimated as per 1st May 2020.
- ** Duration From the date of launch to the date of last contact/issue encounter.
- *** Duration from the date of launch to the date of decay/lost.
- ~~Strikethrough~~ missions are the eliminated data for precise life span prediction.

Bolded texts are the operational mission and their duration is taken as per 1st May 2020.